# Embracing Disorder in Quantum Materials Design

A.R. Mazza[1,*], J. Yan[2], S. Middey[3], J. S. Gardner[2], A.-H. Chen[2], M. Brahlek[2,&], T.Z. Ward[2,+]

[1]Materials Science and Technology Division, Los Alamos National Laboratory, Los Alamos, New Mexico 87545, USA

[2]Materials Science and Technology Division, Oak Ridge National Laboratory, Oak Ridge, TN, 37831, USA

[3] Department of Physics, Indian Institute of Science, Bengaluru 560012, India

[*]ARMazza@lanl.gov, [&]BrahlekM@ornl.gov, [+]WardTZ@ornl.gov

**Abstract:** Many of the most exciting materials discoveries in fundamental condensed matter physics are made in systems hosting some degree of intrinsic disorder. While disorder has historically been regarded as something to be avoided in materials design, it is often of central importance to correlated and quantum materials. This is largely driven by the conceptual and theoretical ease to handle, predict, and understand highly uniform systems that exhibit complex interactions, symmetries and band structures. In this perspective, we highlight how flipping this paradigm has enabled exciting possibilities in the emerging field of high entropy oxide (HEO) quantum materials. These materials host high levels of cation or anion compositional disorder while maintaining unexpectedly uniform single crystal lattices. The diversity of atomic scale interactions of spin, charge, orbital, and lattice degrees of freedom are found to emerge into coherent properties on much larger length scales. Thus, altering the variance and magnitudes of the atomic scale properties through elemental selection can open new routes to tune global correlated phases such as magnetism, metal-insulator transitions, ferroelectricity, and even emergent topological responses. The strategy of embracing disorder in this way provides a much broader pallet from which functional states can be designed for next-generation microelectronic and quantum information systems.

## 1. Introduction

All crystalline materials of any significant size have some degree of disorder. This is a simple result of the large number of constituent atoms, thermodynamics of formation, and synthesis time constraints on the minimization of free energy. Point defects, local atomic displacements, antisite defects, dislocations, and vacancies are amongst the types of extrinsic disorder naturally occurring in crystals. Disorder is ubiquitously present and often has a strong impact on the vast expanse of quantum phenomena. Designing a material around this understanding is a central path to functional control. For example, introducing site-to-site changes to charge state as with substitutional doping. In quantum materials such as high temperature superconductivity, substitutional doping (~10-25%) can be used to used to collapsed Mott insulating and antiferromagnetic ground states of the parent with the addition of charge [1]. Intentional changes to site-to-site atomic radii can also be of practical use. Functional ferroelectricity often arises at the boundary of structural transitions which are readily accessed in alloyed piezo- and relaxor ferroelectrics such as $Ba(Ti_{0.8}Zr_{0.2})O_3$-$x(Ba_{0.7}Ca_{0.3})TiO_3$ (BZT-xBCT), where near the morphotropic phase boundary a large piezoelectric response is driven by local frustration giving rise to polar nanoregions [2,3]. Control of local parameter variation is widely used as a tuning parameter in topological materials (topological insulators, Dirac and Weyl semimetals), where topological properties intrinsically arise in compositions with heavy elements with similar covalencies, such as Bi, Sb, Te, and Se. This chemistry makes them especially prone to high degrees of charge defects (through vacancies or antisite defects) and where the defects have strong impacts on the properties [4–6]. In the prototypical tetradymite topological system, the prime route to mitigate effects of native charge doping is to balance defects through alloying, for example, $(Bi_{1-x}Sb_x)_2(Se_{1-y}Te_y)_3$, where $x$ and $y$ can both range from 0 to 1 [7]. Despite the high level of disorder, these examples show that extremely novel and useful physics exists despite levels of disorder that approach the atomic density.

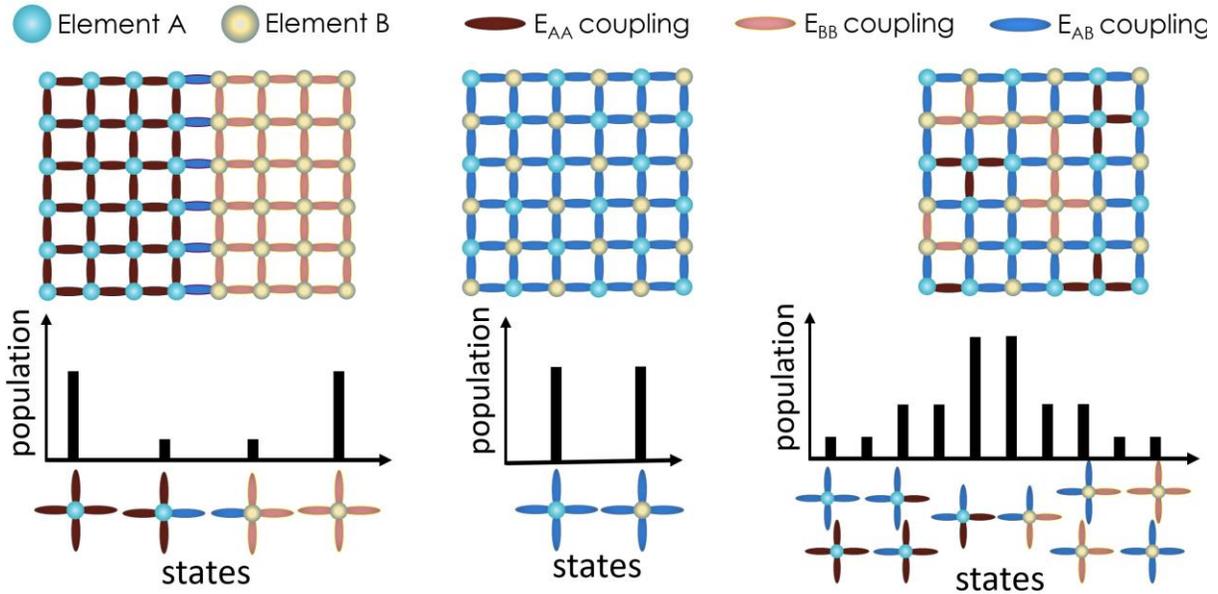

Fig. 1 Diagrams of possible nearest neighbor driven microstates of binary systems. (Left) Heterostructuring creates a small number of microstates that differ from parent materials at the interface. (Center) Ordered systems have regular and repeating local microstates. (Right) Randomly mixed systems host the maximal number of microstates. Due to enthalpic constraints during synthesis, the non-heterostructured examples are often impossible to stabilize. Compositional complexity can increase entropy during synthesis to quench random mixing.

Gaining a more generalized understanding of disorder and learning how it may be exploited in complex many body systems is an outstanding challenge. Is it possible to think of disorder as an order parameter that can be used to design and realize novel physics? A central challenge is in our ability to control and quantify disorder across a much fuller range than simple single element substitutions. Exploration of collective behaviors arising from generalized disorder should be taken in materials that are single phase with randomly disordered well-mixed constituent elements. Counterintuitively, it can be more difficult to create doping maps of well-mixed single-phase materials when there are only a few elements being combined. When substitutionally doping one element for another, the enthalpy of formation has a dominating effect that generally retards mixing of dissimilar elements and leads to phase segregation [8,9].

Towards this, compositionally complex materials (CCMs) have emerged as a platform where disorder can be utilized and explored in very controllable ways while opening nearly infinite disorder phase space. The key idea is to leverage the role entropy plays in the crystallization process, which has long been recognized as a route to drive crystallization through the minimization of the global free energy via configurational entropy [8–11]. As such, increasing disorder through additional elements or site disorder can open routes to stabilize new crystalline materials. This originated in high entropy metals to tune mechanical properties and has since been extended to ionic materials such as oxides. Metal oxides are of specific interests as local chemical disorder impacts the active charge, spin, orbital, and lattice degrees of freedom. These chemical attributes interact on the atomic scale, which then drive magnetic and electronic order on a much larger length scale. This difference in scale then averages over the unit cell level chaos, giving rise to homogeneous responses, much like in the painting technique of pointillism many small, seemingly random dots of paint give rise to beautify images on the macroscale. With strong correlation or in systems which are balanced near metastable or quantum critical points, even a few small variations to the local atomic properties can drive larger changes to the macroscale properties of a material [12,13]. Using materials that have random mixing then permit a much fuller range of local microstates which can lead to unexpected collective macroscopic responses by slight changes to compositional ratios.

Heterostructuring offers a well-travelled path to designing microstates at an interface [14,15] as illustrated in Fig. 1 but offer only very localized and limited affect. Eutectic-like or ordered phases would have a high degree of control over the microstates present, but as was discussed above, enthalpy considerations can make stabilization difficult and sensitive to phase segregation [16]. In a randomly mixed system, there are a much wider range of microstates. With intelligent selection of cations, as example if we can apply our understanding of a specific nearest neighbor interaction to give rise to a targeted magnetic or electronic phase, we may find it possible to design this landscape to favor of an "average" state of choice or to tune toward a quantum critical point that can be driven by miniscule perturbations. For example, selective modification of cations in high entropy perovskite oxides has demonstrated that the macroscopic magnetic state, ordering temperature, and even phase frustration can be "designed" into the system [17].

The premise of this perspective is that complex compositions, high entropy oxides, and related materials [9,18,19] offer an extraordinary and untapped opportunity for discovery of unexpected and highly tunable phenomena in quantum systems. We discuss the role of chemical complexity in controlling electronic and magnetic states in these materials and explore how high entropy materials can address the needs of the materials-physics and chemistry communities. We explore several aspects of CCMs that are particularly relevant to their application: electronic state, topological states, magnetic state, and charge and orbital ordering. We discuss how CCMs can host these states and phenomena, and how they can be manipulated by tuning the composition, disorder, defects, strain, electric field, magnetic field, or (critically) cation variance. In each of these areas, several key questions are addressed in the context of the current experimental and theoretical progress in this field, with the primary focus being on the challenges and opportunities for future research.

## 2. Electronic State

This question is central to both gaining insight into one of the most fundamental quantum mechanical properties of matter and critical to a wide range of applications. In the 1960's P. W. Anderson showed that disorder is sufficient to localize the electron waves packets responsible for conduction in metals [20,21]. For materials with purely metallic bonding (e.g. sodium or copper) the electrons can screen disorder on a unit cell level, thus guaranteeing a metallic state for the most disordered scenarios. In contrast, for covalent and ionic materials such as the high entropy oxides [8,9] the screening length is much longer, thereby making disorder-induced localization effects more pronounced. However, there are observations of both metallic and insulating chemically complex oxide systems. Specifically, the $R$NiO$_3$ ($R$ being ~5 rare-earth elements) [22,23] and Riddlesden-Popper cuprates [24,25] are found to be metallic, whereas La(Cr,Mn,Fe,Co,Ni)O$_3$ [26–29] and the original J14 rocksalts [9,30] are well-known to be strongly insulating. There are several novel observations and important questions regarding these.

First, in the $R$NiO$_3$ system the conduction occurs entirely through the Ni orbitals. The rare earth site is completely empty, and the bonding orbitals are well away from the Fermi level. Therefore, disorder affects the conduction electrons mostly through changing the Ni-O-Ni bond angles, which is sufficiently minimal to enable the preservation of the metallic phase. In the latter cases, the disorder is directly on the states near the Fermi level. Therefore, despite long range crystalline homogeneity [23], the large variations in local energy states are a sufficient barrier to delocalization and conduction. Are these strictly examples of Anderson insulators or is there more complex physics at play?

The two insulating examples [9,27] are systems composed of parents where effects of strong electron-electron correlations induced localization are pronounced and the parents fall into the category of Mott insulators [31]. This motivates several questions where chemically complex materials may be a new playground to answer fundamental questions. Specifically, an area of focus must be in understanding how disorder and electron-correlations compete or conspire to drive materials that should be metallic into an insulating phase. Theory has been developed to sketch disorder-correlation phase diagrams for specific cases [32]. However, mapping such diagrams to experimentally testable parameter spaces is a muddy process. For systems such as the high temperature superconducting cuprates, the Mott-insulating parent is strongly doped, disorder is increased, correlation effects are reduced and band filling increases, creating a complex path through this disorder-correlation space. In contrast, chemical complexity offers a new route to systematically traverse this landscape. For example, for the $R$NiO$_3$, $R$ can be chosen such that there is a large or small variance in the distribution of $R$. This manifests as a geometric change in local bond angles, which may be trackable since correlation effects tend to increase with decreasing bond angle (relative to 180°) [23,33,34]. Disorder strength should scale similarity, yet how disorder should be quantified is still an open question. Deploying now-well-established experimental probes such as angle-resolved photo-emission spectroscopy (APRES), may enable extraction of the renormalized band mass, which has been effective in quantifying correlation effects. Combining this with transport probes will enable distinguishing aspects of disorder and correlations. Altogether, understanding how to utilize this new materials paradigm of chemical complexity will enable looking at materials exploration in a new light.

## 3. Topological States

Topologically protected states are quantum states of matter that are robust against external perturbations and can support exotic phenomena, such as topological insulators [35–37], Weyl semimetals [38], and Majorana fermions [39]. Topology has recently become a central tenant to understand fundamental properties of matter. This ranges from basic effects of band structures to magnetic textures [5,37,38,40,41]. Topology arises as a categorization scheme to distill the broad properties of materials into a single number, the topological invariant. Regarding band structures, this number is derived as an integral over the wavefunctions, and, thus, is based on how the atomic energy bands are ordered. The physical properties manifest at the boundary (edge or surfaces) among non-topological and topological materials where unusual metallic states arise and must exist so long as the band structure is intact. In the light of chemically complex materials, several important questions arise.

Are band structures well-defined in CCMs, and, if so, can a topological HEO be found? This question ploughs deeply into whether periodic translational symmetry, which is a key requirement for well-defined energy-moment relation, of a lattice is fundamentally valid. For a perfect crystalline material, the potential landscape is periodic, and electrons placed into this landscape will form Bloch waves with dispersing bands with well-separated gaps [42]. Alternatively, constructing a lattice from isolated chemical bonds, dispersive bands can be built from all the possible patterns of bonding and antibonding orbitals, each of which specifies an explicit point of the energy-moment relation [43]. In either viewpoint, the notion of band structure, and thus topological aspects, fundamentally fails for the case of chemical complex materials since periodic transitional symmetry is lost. However, it may be rescued if the atomic-site perturbations are small. Taking the case of the $R$NiO$_3$, again, the disorder is on the $R$ site, whose 5s atomic orbitals are empty and well above the Fermi level. Therefore, the Ni bonding geometry are only slightly perturbed, and, therefore, a band structure may form with properties slightly perturbed. In the case of La(Cr,Mn,Fe,Co,Ni)O$_3$ the bonding orbitals near the Fermi level are strongly disordered site-to-site, have different fillings and character, and, therefore, it is likely that the notion of well-formed bands gives way to isolated energy levels, more akin to organic semiconductors where conducting and valence bands are replaced by highest-occupied molecular orbitals (HOMO) and lowest unoccupied molecular orbitals (LUMO) levels. As such, the notion of topology for chemical complexity may arise in specific systems where the bonds near the Fermi level are derived from the non-mixed site, or in unexpected situations where topology no longer requires well-defined band structures as a defining characteristic [44]. Here, when the topological insulator was made amorphous it becomes insulating, but initial ARPES indicates it may remain topological [44–46]. This prompts several questions regarding the nature of this disorder induced transitions, and whether it follows the nature of topological phase transitions driven solely by weakening of spin orbit coupling and finite size effects [47–50]. Conversely, does disorder drive a new type of topological-to-topological transition? An interesting example is the magnetic topological insulator $MnBi_2Te_4$, where the topological bands derive from Bi and Te orbitals. The magnetism arises from the Mn site, whose bonding orbitals are well away from the Fermi level. This may be an ideal system to explore how adding site complexity to the Mn position in the form of size-balanced transition metal cations with valence 2+ may enable overcoming synthesis challenges [51,52] while addressing fundamental questions.

Can chemical complexity be used to improve properties of topological materials to boost the useful temperature towards room temperature? The novel aspects of topological states have made them a theorist's playground, which have resulted in a myriad of predicted technologies [53]. However, materials issues have been the limiting factor to bridge basic properties with realistic applications. The specific problem is that the band gaps of known topological materials are all small (≤0.3eV), the static dielectric constants are large (≥100), the typically effective masses are small (0.1-0.2× bare electron mass), and they have a large density of native defects [5]. These deleterious effects are fundamentally rooted in the necessary ingredients required for a material to be topological [54]. Specifically, the necessary band inversion, driven by strong spin-orbital coupling, requires constituent elements to be heavy for strong spin-orbital coupling, and chemically similar to create sufficiently narrow the spin-orbit to invert the bands. The best materials are therefore prone to chemical defects, and, even with optimized defect control, have sufficiently small band gaps to necessitate cryogenic temperatures to realize novel properties. As such, if topology is robust to strong levels of disorder, utilizing the tools of chemical complexity may offer unexpected routes to overcome some of these basic challenges. Specifically, selectively choosing cations and anions in the $Bi_2Se_3$/$Bi_2Te_3$ tetradymites has been effective at positioning the

Fermi level [55]. Going to the chemically complex limit may offer greater control via changing the defect landscape, tuning the character of the bands by tuning the effective mass and possible overcoming of the 0.3 eV band gap barrier, or by pushing to lighter elements where the static dielectric constant can be effectively reduced. Modest improves to these critical parameters, for example, even a modest 0.1 eV improvement in band gap or 50% reduction in static dielectric constant would result in materials that are much more robust to defects. This would allow access to topological states at room temperature or above, which is a critical step towards translating these behaviors to practical topological technologies.

**4. Magnetic State**

Magnetic frustration has largely been studied in the light of geometrical frustration – where the magnetic ground state is degenerate due to the geometry of the lattice. However, recent work in high entropy oxides has demonstrated configurational complexity can give rise to frustration driven by local disorder in spin and exchange interactions [17,56]. The spatial correlations of spins within a material are a result of complex exchange interactions, often dominated by nearest neighbors, but in reality, further neighbor couplings must be appreciated to understand the exact nature of the arrangement of spins and their dynamics. For example, in $Ca_{10}Cr_7O_{28}$, 5 interactions were needed to understand the nature of the spin system [57,58]. However, a cubic perovskite HEO system with 5 different elements at magnetically active and coordinated sites would have a minimum of 15 unique exchange interaction possibilities on the microstate even if one neglects interactions beyond the first nearest neighbor and any local distortions [12]. Continued research in this area is a promising avenue towards searching for frustrated states, with the following questions driving the understanding of dynamics in configurationally complex materials.

Can the understanding of magnetic phase and phase degeneracy in high entropy oxides extend to spin-liquid, ice, or glass like behaviors? Early work towards understanding of HEO magnetism focused on the prevailing order type, with the intriguing result of long range magnetic order in such a chaotic chemical environment being a surprise to the community [17,59,60]. However, in these works and subsequent studies of HEOs the prevailing magnetic order type was clearly not the complete story. A tendency to observe gradual magnetic transitions rather than abrupt ordering as seen in the simple oxides and an indication that the ordered fraction never reaches 100% for some HEOs were clear signs of frustration. Efforts to understand dynamics have shown that disorder in these systems drives fluctuations near and in some cases below the magnetic order temperature. To date, most of the HEO spin glass materials are a matrix of frustrated AFM and spin glass regions – often giving rise to biased pinning of moments which exhibit tunability and broad control of reversal [17,61,62]. Exploring these phase frustrated behaviors in HEOs and their role in technologically relevant phenomena like exchange bias promises an interesting path forward in exploring materials for future spin-based electronics.

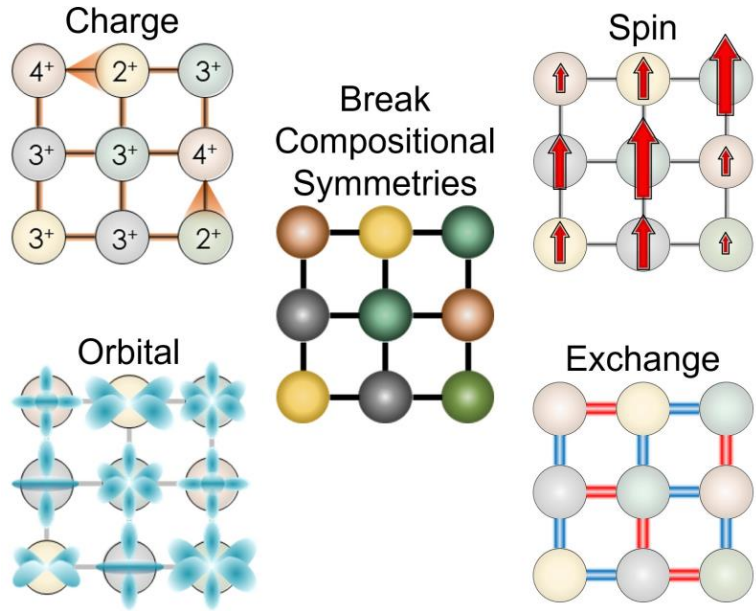

Fig. 2 Random mixing of composition on a crystal lattice provides a route to conserve long range structural symmetries while breaking charge, spin, orbital, and magnetic exchange interaction symmetries across all length scales.

In Fig. 2 the relationship between compositional symmetry breaking and resulting spin and exchange disorder is illustrated. Clearly, in systems with a variety of spin states and exchange interactions, a degenerate or frustrated magnetic ground state could be realized by design. Configurational complexity has already been shown to drive dynamical and phase frustrated behaviors in perovskite, rocksalt, pyrochlore, and spinel oxides [56,61,63–66]. Certainly, the broad tunability of the prevailing magnetic state in HEOs suggests the degeneracy can be controlled, however pushing this into a state in which no prevailing magnetic order arises is complex. Access to quantum critical points is certainly possible [17]. Spin-liquid like behaviors do not appear outside the realm of possibilities, particularly in considering theoretical work showing that randomization of exchange interactions on a square lattice can lead to such a state [67–69]. Experimentally, exchange disorder in symmetric systems (i.e. square lattices with S = ½) has been shown to lead to a spin-liquid like state [70,71]. This work suggests careful consideration and construction of the spin, charge, and exchange interactions between constituent elements in a candidate CCM system could yield liquid-like behavior.

Are we limited to known phase Are we limited to known phase diagrams (i.e. the magnetic exchange interactions of the parent oxides) in controlling magnetic phase? The broadly tunable nature of the magnetic state in high entropy oxides has been demonstrated – where an incipient critical point between prevailing AFM and FM phases was mapped in high entropy transition metal perovskites [17]. Atomic scale order has been seen to extend to macroscopic spin textures in single crystal high entropy spinel oxide films [72], where the size and shape of macroscopic magnetic stripes is dependent on heteroepitaxial strain. This tunability appears to extend into other high entropy oxides [73,74], though it has not been explored towards understanding how variance, much like in the case of the electronic state of nickelates [23], might impact and promote magnetic ordering temperatures which exceed that of the parent oxides. Not surprisingly, the most prevalent change observed in the magnetism of high entropy oxides is the glassiness of transitions as compared to simpler systems [56,62]. However, given the nontrivial coupling of lattice, charge and spin order in oxides randomization of not only the spin but of the local oxygen octahedral rotations [75] with a large cation size variance could lead to local opening of double or superexchange pathways not available in the parent compounds. Furthermore, the continuous tunability of charge and its effect on magnetism is just beginning to be explored – where already the effect of hole doping in perovskite systems shows control of local double exchange pathways in transition metal perovskite HEOs [26,61]. Continuing to explore local charge [76,77] and how to manipulate it is imperative to understanding how far magnetic phase can be pushed outside of understood phase diagrams.

## 5. Charge and Orbital Order

Many of the intriguing properties of strongly correlated oxide systems are closely tied to orbital and charge degrees of freedom in these systems. The nature of magnetic exchange and the resulting magnetic structure of complex oxides, for example, depends on the orbital character (and oxidation state) of the valence electrons of transition metal ions. In the orbital degenerate system, the degeneracy is lifted by structural distortions, leading to orbital ordering (OO) [78]. Interestingly, OO can also happen due to electronic superexchange, known as the Kugel-Khomskii mechanism [79]. This relationship is especially evident in rare-earth manganites. For example, properties such as magnetism, colossal magnetoresistance, and metal to insulator transitions are driven by changes in Mn valence and orbital ordering [80–82]. Given this context, extreme disorder at the rare earth site becomes a highly motivating tuning knob in exploring functional phase space when considering this connection, with preliminary results reporting that a high degree of rare-earth size variance drives significant effects on charge and orbital ordering in

HEOs [23,83–87]. These early results show two key areas of promise. First, high entropy synthesis can be applied to deepen our understanding of the physics of extreme disorder. Second, using this type of disorder to move beyond established phase diagrams (or the average state) to novel ordered states is possible. This brings about several key questions.

Can charge/orbital ordering survive extreme disorder? The question of whether or not orbital order in CCMs prevails is two-fold. First, in the case in which the functional site is not disordered chemically. This issue has been investigated for rare-earth vanadates ($R$VO$_3$). Depending on the choice of $R$ cation, this family exhibits G-type OO (with C-type antiferromagnetic ordering) and C-type OO (with G-type antiferromagnetism) [83]. The notable observation is that the average dictates the ordering except in the case of a high degree of size variance of ionic radius of the RE, which act to stabilize/suppress OO in the HEO. Studying this aspect for the $R$TiO$_3$ family would be interesting in the same vein, where the role of orbital physics for setting the antiferromagnetism has remained a subject of intense scrutiny as the Kugel-Khomskii mechanism predicts ferromagnetism in the $t_{2g}^1$ system [88]. Orbital physics also strongly influences phenomena like metal-insulator transitions, colossal magnetoresistance, and unconventional superconductivity [89–91]. Therefore, extending the family of materials in which these studies determine the role of extreme disorder in suppressing/enhancing competing order types is an important next step. The second, more complex issue is to address the question of OO of materials with the chemically complex site being the functional site (i.e. disorder on the $B$-site of an $AB$O$_3$ perovskite). This is again illustrated in Fig. 2 where compositional symmetry breaking leads to a complex landscape in terms of the site-to-site orbital and charge populations. Can global orbital or charge order occur on a compositionally disorder sublattice? Little work has been done in this direction but, given the observation of long-range magnetic order, local and even global OO appears plausible.

Many transition metal (TM) compounds with a non-integer number of $d$ electrons per site exhibit charge ordering (CO) [78]. In such systems, the high-temperature phase has equivalent TM sites or random TM sites with a variation of oxidation state, while the low-temperature CO insulating phase has a periodic structure with inequivalent valence states. Such CO leads to metal-insulator transitions, semiconductor to insulator transitions, and magnetic transitions [92–94]. A particular case is CE-type charge orbital ordering in half-doped manganite, such as charge, orbital, and spin ordering in Nd$_{0.5}$Sr$_{0.5}$MnO$_3$ [95,96]. The fate of these transitions in the presence of strong disorder remains an open question, though HEO manganites are a rapidly expanding field sure to provide important new insights [84,85,97–100]. CO is also considered the driving mechanism of the metal-insulator transition of bulk $RNiO_3$ [101]. However, considering the importance of ligand hole states, the MIT is described presently as a bond disproportionation (BD) transition, which can be considered a 'charge ordering' on the oxygen sublattice [34,102,103]. This has been explored in HEO nickelates [22,23], where the BD/CO transition is modified via cation size variations.

How can a large cation size variance impact charge/orbital ordering? Perovskite oxides are an excellent candidate to explore this question. Doping the rare earth site on systems like $(RE)VO_3$ and $(RE)MnO_3$ is well established and largely connected to local lattice distortions. Examining extreme disorder – with an eye set on variance in the size of the cation at the compositionally complex rare earth size – begins to demonstrate how frustration on the nanoscale can impact orbital and charge ordering at meso- and macroscales. Local distortions are known to be induced by site-to-site variance in cation size in HEOs [23,75], and that connection seems to drive divergence from expected collective states [23,83]. Perhaps the most compelling exploration of the role of size variance is in the context of orbital ordering in $(RE)VO_3$, which has shown that when the size variance at the perovskite RE-site is small, spin and orbital ordering temperatures fit well to that expected for a single average RE size. As variance increased for a given average, orbital order is impacted, and the competing spin-orbital orderings become intertwined [83]. Similarly, large size variance in the $(RE)NiO_3$ system destabilizes charge disproportionation due to induced frustration in the Ni-O-Ni bond angle, which decouples the electronic and structural phase transition [23]. These examples merely begin to demonstrate how variance can be used to tune, beyond known phase diagrams, spin, charge, and orbital degrees of freedom in compositionally complex systems.

How does hole/electron doping of the disordered site impact functionality? Understanding of local charge compensation mechanisms in HEOs remain elusive. Studies have observed mixed valence at like sites [76,77], however it is unclear whether this charge compensation happens locally or over several nearest neighbors. Nor is it clear how much valence flexibility is accessible through compositional tuning. In principle, considering Pauling's 2$^{nd}$ rule of electrostatic valence, neutralizing of valence should be a simple process including only the first nearest neighbor. However, in the complex environment of CCMs, neutralization may not be so simple. The local distortions, competing chemistries, variance in electronegativity from site to site can greatly impact the static valance of an element in these solid-state solutions. Hole doping in HEO transition metal perovskites seems to hint at a non-linear change in valence. Here, double exchange leading to FM pockets in the material show that local 4+ sites compensate 2+ sites rather than a commensurate shift where each site compensates equally with the introduction of holes to the system [61]. Exploring directly how valence is impacted by hole doping and the electron dynamics/localization in such systems will begin to lend first of its kind understanding to the question of how charge compensation occurs in the limit of extreme disorder.

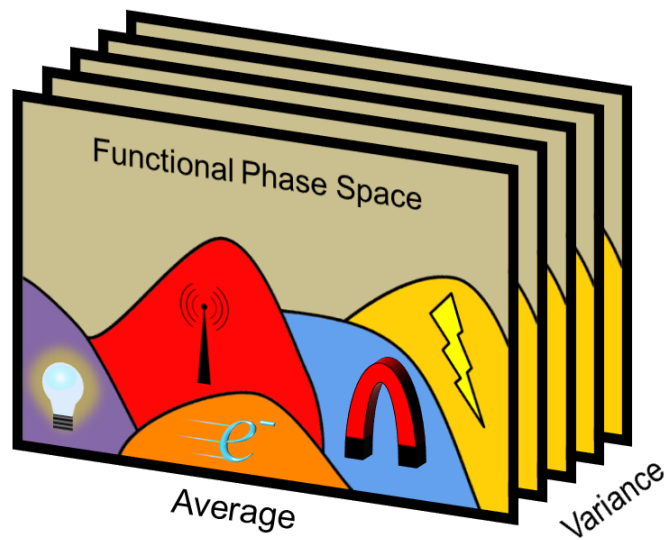

Fig. 3 A representation of the many dimensions to explore in functional phase space. The "average" state of the spin, charge, orbital, and lattice degrees of freedom can be continuously tuned to a desired state. The variance in these degrees of freedom is seen as a promising avenue to break outside of known physics and average functional behaviors in CCMs.

## 6. Summary and Conclusions

Compositional complexity demonstrates a novel avenue toward quantum materials design. HEOs and newer systems like high entropy chalcogenides [104,105] are at the center of this strategy, displaying unique control over quantum phenomena with composition and degree of disorder (variance). The continuous tuneability of composition in HEOs (see Fig. 3) gives unsurpassable access to functional phase space. The simplest physics when considering these materials is that, with sufficiently small variance, the average state can be continuously tuned to the desired functionality. However, we suggest a new order parameter, beyond that of elemental disorder, which has emerged in the form of cation variance. This adds a dimension to explore in phase space (Fig. 3), where HEOs have been shown to diverge from expected behaviors in nickelate and vanadate perovskites [23,83]. We have discussed variance in the context of cation size, but the spin (and magnetic exchange), charge (i.e. aliovalent dopants), and orbital degrees of freedom can be explored in much the same way. These areas express some of the most exciting areas in which novel physics, largely at the limits of extreme disorder, can be explored. There have already been surprises in this area, particularly in the context of magnetism, where the average state produces long range order while frustration drives novel interactions such as monolithic exchange bias [17]. Several other materials, from ferroelectrics [106,107] to superconductors [24,108,109] are emerging in unexpected ways. This nascent field is rapidly maturing and the future of exploring disorder and variance as a parameter towards accessing novel functional phases appears bright amongst CCMs where harnessing disorder for exploitation of novel functionality offers important new directions in future microelectronic and quantum information applications.

**Acknowledgements**

This work was supported by the U.S. Department of Energy (DOE), Office of Science, Basic Energy Sciences (BES), Materials Sciences and Engineering Division and the NNSA's Laboratory Directed Research and Development Program at Los Alamos National Laboratory. Los Alamos National Laboratory, an affirmative action equal opportunity employer, is managed by Triad National Security, LLC for the U.S. Department of Energy's NNSA, under contract 89233218CNA000001. SM is funded by Quantum Research Park, which is a project administered by FSID, IISc with support from KITS, Government of Karnataka.

**References**

[1] P. A. Lee, N. Nagaosa, and X.-G. Wen, *Doping a Mott Insulator: Physics of High-Temperature Superconductivity*, Rev. Mod. Phys. **78**, 17 (2006).
[2] N. Cucciniello, A. R. Mazza, P. Roy, S. Kunwar, D. Zhang, H. Y. Feng, K. Arsky, A. Chen, and Q. Jia, *Anisotropic Properties of Epitaxial Ferroelectric Lead-Free 0.5[Ba(Ti0.8Zr0.2)O3]-0.5(Ba0.7Ca0.3)TiO3 Films*, Materials **16**, 6671 (2023).
[3] R. Yuan et al., *Machine Learning-Enabled Superior Energy Storage in Ferroelectric Films with a Slush-Like Polar State*, Nano Lett. **23**, 4807 (2023).
[4] M. Brahlek, N. Koirala, N. Bansal, and S. Oh, *Transport Properties of Topological Insulators: Band Bending, Bulk Metal-to-Insulator Transition, and Weak Anti-Localization*, Solid State Commun. **215–216**, 54 (2015).
[5] M. Brahlek, J. Lapano, and J. S. Lee, *Topological Materials by Molecular Beam Epitaxy*, J. Appl. Phys. **128**, 210902 (2020).
[6] M. Salehi, X. Yao, and S. Oh, *From Classical to Quantum Regime of Topological Surface States via Defect Engineering*, SciPost Phys. Lect. Notes 58 (2022).
[7] J. Zhang et al., *Band Structure Engineering in (Bi1−xSbx)2Te3 Ternary Topological Insulators*, Nat. Commun. **2**, 574 (2011).
[8] M. Brahlek et al., *What Is in a Name: Defining "High Entropy" Oxides*, APL Mater. **10**, 110902 (2022).
[9] C. M. Rost, E. Sachet, T. Borman, A. Moballegh, E. C. Dickey, D. Hou, J. L. Jones, S. Curtarolo, and J.-P. Maria, *Entropy-Stabilized Oxides*, Nat. Commun. **6**, 8485 (2015).
[10] G. N. Kotsonis, *Property and Cation Valence Engineering in Entropy-Stabilized Oxide Thin Films*, Phys. Rev. Mater. 9 (2020).

[11] D. B. Miracle and O. N. Senkov, *A Critical Review of High Entropy Alloys and Related Concepts*, Acta Mater. **122**, 448 (2017).

[12] H. Guo, J. H. Noh, S. Dong, P. D. Rack, Z. Gai, X. Xu, E. Dagotto, J. Shen, and T. Z. Ward, *Electrophoretic-like Gating Used To Control Metal–Insulator Transitions in Electronically Phase Separated Manganite Wires*, Nano Lett. **13**, 3749 (2013).

[13] E. Skoropata, A. R. Mazza, A. Herklotz, J. M. Ok, G. Eres, M. Brahlek, T. R. Charlton, H. N. Lee, and T. Z. Ward, *Post-Synthesis Control of Berry Phase Driven Magnetotransport in SrRuO 3 Films*, Phys. Rev. B **103**, 085121 (2021).

[14] T. Z. Ward, Z. Gai, X. Y. Xu, H. W. Guo, L. F. Yin, and J. Shen, *Tuning the Metal-Insulator Transition in Manganite Films through Surface Exchange Coupling with Magnetic Nanodots*, Phys. Rev. Lett. **106**, 157207 (2011).

[15] Z. Gai et al., *Chemically Induced Jahn–Teller Orderings on Manganite Surfaces*, Nat. Commun. **5**, 4528 (2014).

[16] G. Cao et al., *Ferromagnetism and Nonmetallic Transport of Thin-Film $\alpha - FeSi_2$: A Stabilized Metastable Material*, Phys. Rev. Lett. **114**, 147202 (2015).

[17] A. R. Mazza et al., *Designing Magnetism in High Entropy Oxides*, Adv. Sci. **9**, 2200391 (2022).

[18] P. B. Meisenheimer and J. T. Heron, *Oxides and the High Entropy Regime: A New Mix for Engineering Physical Properties*, MRS Adv. **5**, 3419 (2020).

[19] Y. Sharma et al., *Single-Crystal High Entropy Perovskite Oxide Epitaxial Films*, Phys. Rev. Mater. **2**, 060404 (2018).

[20] P. W. Anderson, *Absence of Diffusion in Certain Random Lattices*, Phys. Rev. **109**, 1492 (1958).

[21] C. Kittel, *Introduction to Solid State Physics*, 8th ed. (Wiley, 2004).

[22] R. K. Patel, S. K. Ojha, S. Kumar, A. Saha, P. Mandal, J. W. Freeland, and S. Middey, *Epitaxial Stabilization of Ultra Thin Films of High Entropy Perovskite*, Appl. Phys. Lett. **116**, 071601 (2020).

[23] A. R. Mazza et al., *Variance Induced Decoupling of Spin, Lattice, and Charge Ordering in Perovskite Nickelates*, Phys. Rev. Res. **5**, 013008 (2023).

[24] A. R. Mazza et al., *Searching for Superconductivity in High Entropy Oxide Ruddlesden–Popper Cuprate Films*, J. Vac. Sci. Technol. A **40**, 013404 (2022).

[25] W. Zhang et al., *Applying Configurational Complexity to the 2D Ruddlesden–Popper Crystal Structure*, ACS Nano **14**, 13030 (2020).

[26] A. R. Mazza et al., *Charge Doping Effects on Magnetic Properties of Single-Crystal $La_{1-x}Sr_x(Cr_{0.2}Mn_{0.2}Fe_{0.2}Co_{0.2}Ni_{0.2})O_3$ ($0 \leq x \leq 0.5$) High-Entropy Perovskite Oxides*, Phys. Rev. B **104**, 094204 (2021).

[27] Y. Sharma et al., *Magnetic Anisotropy in Single-Crystal High-Entropy Perovskite Oxide $La(Cr_{0.2}Mn_{0.2}Fe_{0.2}Co_{0.2}Ni_{0.2})O_3$ Films*, Phys. Rev. Mater. **4**, 014404 (2020).

[28] A. Sarkar, R. Djenadic, D. Wang, C. Hein, R. Kautenburger, O. Clemens, and H. Hahn, *Rare Earth and Transition Metal Based Entropy Stabilised Perovskite Type Oxides*, J. Eur. Ceram. Soc. **38**, 2318 (2018).

[29] M. Brahlek, A. R. Mazza, K. C. Pitike, E. Skoropata, J. Lapano, G. Eres, V. R. Cooper, and T. Z. Ward, *Unexpected Crystalline Homogeneity from the Disordered Bond Network in $La(Cr_{0.2}Mn_{0.2}Fe_{0.2}Co_{0.2}Ni_{0.2})O_3$ Films*, Phys. Rev. Mater. **4**, 054407 (2020).

[30] C. M. Rost, Z. Rak, D. W. Brenner, and J.-P. Maria, *Local Structure of the $Mg_x Ni_x Co_x Cu_x Zn_x O$ ($x=0.2$) Entropy-Stabilized Oxide: An EXAFS Study*, J. Am. Ceram. Soc. **100**, 2732 (2017).

[31] N. F. Mott, *The Basis of the Electron Theory of Metals, with Special Reference to the Transition Metals*, Proc. Phys. Soc. Sect. A **62**, 416 (1949).

[32] K. Byczuk, W. Hofstetter, and D. Vollhardt, *Mott-Hubbard Transition versus Anderson Localization in Correlated Electron Systems with Disorder*, Phys. Rev. Lett. **94**, 056404 (2005).

[33] S. Catalano, M. Gibert, J. Fowlie, J. Íñiguez, J.-M. Triscone, and J. Kreisel, *Rare-Earth Nickelates R NiO$_3$ : Thin Films and Heterostructures*, Rep. Prog. Phys. **81**, 046501 (2018).

[34] S. Middey, J. Chakhalian, P. Mahadevan, J. W. Freeland, A. J. Millis, and D. D. Sarma, *Physics of Ultrathin Films and Heterostructures of Rare-Earth Nickelates*, Annu. Rev. Mater. Res. **46**, 305 (2016).

[35] J. Lapano et al., *Strong Spin-Dephasing in a Topological Insulator-Paramagnet Heterostructure*, APL Mater. **8**, 091113 (2020).

[36] Y. Xia et al., *Observation of a Large-Gap Topological-Insulator Class with a Single Dirac Cone on the Surface*, Nat. Phys. **5**, 398 (2009).

[37] Y. Deng, Y. Yu, M. Z. Shi, Z. Guo, Z. Xu, J. Wang, X. H. Chen, and Y. Zhang, *Quantum Anomalous Hall Effect in Intrinsic Magnetic Topological Insulator MnBi$_2$Te$_4$*, Science **367**, 895 (2020).

[38] B. Yan and C. Felser, *Topological Materials: Weyl Semimetals*, Annu. Rev. Condens. Matter Phys. **8**, 337 (2017).

[39] C. W. J. Beenakker, *Search for Majorana Fermions in Superconductors*, Annu. Rev. Condens. Matter Phys. **4**, 113 (2013).

[40] X.-L. Qi and S.-C. Zhang, *Topological Insulators and Superconductors*, Rev. Mod. Phys. **83**, 1057 (2011).

[41] O. Gomonay, V. Baltz, A. Brataas, and Y. Tserkovnyak, *Antiferromagnetic Spin Textures and Dynamics*, Nat. Phys. **14**, 213 (2018).

[42] N. W. Ashcroft and N. D. Mermin, *Solid State Physics* (Saunders college publ, Fort Worth Philadelphia San Diego [Etc.], 1976).

[43] R. Hoffmann, *How Chemistry and Physics Meet in the Solid State*, Angew. Chem. Int. Ed. Engl. **26**, 846 (1987).

[44] M. Brahlek et al., *Disorder-Driven Topological Phase Transition in $Bi_2Se_3$ Films*, Phys. Rev. B **94**, 165104 (2016).

[45] A. Agarwala and V. B. Shenoy, *Topological Insulators in Amorphous Systems*, Phys. Rev. Lett. **118**, 236402 (2017).

[46] P. Corbae et al., *Observation of Spin-Momentum Locked Surface States in Amorphous Bi2Se3*, Nat. Mater. **22**, 200 (2023).

[47] M. Brahlek, N. Bansal, N. Koirala, S.-Y. Xu, M. Neupane, C. Liu, M. Z. Hasan, and S. Oh, *Topological-Metal to Band-Insulator Transition in $(Bi_{1-x}In_x)_2Se_3$ Thin Films*, Phys. Rev. Lett. **109**, 186403 (2012).

[48] L. Wu, M. Brahlek, R. Valdés Aguilar, A. V. Stier, C. M. Morris, Y. Lubashevsky, L. S. Bilbro, N. Bansal, S. Oh, and N. P. Armitage, *A Sudden Collapse in the Transport Lifetime across the Topological Phase Transition in (Bi1−xInx)2Se3*, Nat. Phys. **9**, 410 (2013).

[49] M. J. Brahlek, N. Koirala, J. Liu, T. I. Yusufaly, M. Salehi, M.-G. Han, Y. Zhu, D. Vanderbilt, and S. Oh, *Tunable Inverse Topological Heterostructure Utilizing $(Bi_{1-x}In_x)_2Se_3$ and Multichannel Weak-Antilocalization Effect*, Phys. Rev. B **93**, 125416 (2016).

[50] M. Salehi, H. Shapourian, N. Koirala, M. J. Brahlek, J. Moon, and S. Oh, *Finite-Size and Composition-Driven Topological Phase Transition in $(Bi_{1-x}In_x)_2Se_3$ Thin Films*, Nano Lett. **16**, 5528 (2016).

[51] A. R. Mazza et al., *Surface-Driven Evolution of the Anomalous Hall Effect in Magnetic Topological Insulator MnBi$_2$Te$_4$ Thin Films*, Adv. Funct. Mater. **32**, 2202234 (2022).

[52] J. Lapano et al., *Adsorption-Controlled Growth of MnTe(Bi$_2$Te$_3$)n by Molecular Beam Epitaxy Exhibiting Stoichiometry-Controlled Magnetism*, Phys. Rev. Mater. **4**, 111201 (2020).

[53] M. J. Gilbert, *Topological Electronics*, Commun. Phys. **4**, 70 (2021).

[54] M. Brahlek, *Criteria for Realizing Room-Temperature Electrical Transport Applications of Topological Materials*, Adv. Mater. **32**, 2005698 (2020).

[55] R. Jiang, L.-L. Wang, M. Huang, R. S. Dhaka, D. D. Johnson, T. A. Lograsso, and A. Kaminski, *Reversible Tuning of the Surface State in a Pseudobinary Bi 2 (Te-Se) 3 Topological Insulator*, Phys. Rev. B **86**, 085112 (2012).

[56] P. B. Meisenheimer et al., *Magnetic Frustration Control through Tunable Stereochemically Driven Disorder in Entropy-Stabilized Oxides*, Phys. Rev. Mater. **3**, 104420 (2019).

[57] C. Balz, B. Lake, M. Reehuis, A. T. M. Nazmul Islam, O. Prokhnenko, Y. Singh, P. Pattison, and S. Tóth, *Crystal Growth, Structure and Magnetic Properties of Ca $_{10}$ Cr $_7$ O $_{28}$*, J. Phys. Condens. Matter **29**, 225802 (2017).

[58] C. Balz et al., *Physical Realization of a Quantum Spin Liquid Based on a Complex Frustration Mechanism*, Nat. Phys. **12**, 942 (2016).

[59] J. Zhang et al., *Long-Range Antiferromagnetic Order in a Rocksalt High Entropy Oxide*, Chem. Mater. **31**, 3705 (2019).

[60] Zs. Rák and D. W. Brenner, *Exchange Interactions and Long-Range Magnetic Order in the (Mg,Co,Cu,Ni,Zn)O Entropy-Stabilized Oxide: A Theoretical Investigation*, J. Appl. Phys. **127**, 185108 (2020).

[61] A. R. Mazza et al., *Hole Doping in Compositionally Complex Correlated Oxide Enables Tunable Exchange Biasing*, APL Mater. **11**, 031118 (2023).

[62] P. B. Meisenheimer, T. J. Kratofil, and J. T. Heron, *Giant Enhancement of Exchange Coupling in Entropy-Stabilized Oxide Heterostructures*, Sci. Rep. **7**, 13344 (2017).

[63] S. Marik, D. Singh, B. Gonano, F. Veillon, D. Pelloquin, and Y. Bréard, *Enhanced Magnetic Frustration in a New High Entropy Diamond Lattice Spinel Oxide*, Scr. Mater. **186**, 366 (2020).

[64] B. Jiang, C. A. Bridges, R. R. Unocic, K. C. Pitike, V. R. Cooper, Y. Zhang, D.-Y. Lin, and K. Page, *Probing the Local Site Disorder and Distortion in Pyrochlore High-Entropy Oxides*, J. Am. Chem. Soc. **143**, 4193 (2021).

[65] D. Karoblis, O. C. Stewart, P. Glaser, S. E. El Jamal, A. Kizalaite, T. Murauskas, A. Zarkov, A. Kareiva, and S. L. Stoll, *Low-Temperature Synthesis of Magnetic Pyrochlores (R $_2$ Mn $_2$ O $_7$, R = Y, Ho–Lu) at Ambient Pressure and Potential for High-Entropy Oxide Synthesis*, Inorg. Chem. **62**, 10635 (2023).

[66] P. R. Jothi, W. Liyanage, B. Jiang, S. Paladugu, D. Olds, D. A. Gilbert, and K. Page, *Persistent Structure and Frustrated Magnetism in High Entropy Rare-Earth Zirconates*, Small **18**, 2101323 (2022).

[67] J. Vannimenus and G. Toulouse, *Theory of the Frustration Effect. II. Ising Spins on a Square Lattice*, J. Phys. C Solid State Phys. **10**, L537 (1977).

[68] N. Laflorencie, S. Wessel, A. Läuchli, and H. Rieger, *Random-Exchange Quantum Heisenberg Antiferromagnets on a Square Lattice*, Phys. Rev. B **73**, 060403 (2006).

[69] P. Chandra and B. Doucot, *Possible Spin-Liquid State at Large S for the Frustrated Square Heisenberg Lattice*, Phys. Rev. B **38**, 9335 (1988).

[70] O. Mustonen, S. Vasala, K. P. Schmidt, E. Sadrollahi, H. C. Walker, I. Terasaki, F. J. Litterst, E. Baggio-Saitovitch, and M. Karppinen, *Tuning the S = 1 / 2 Square-Lattice Antiferromagnet S r 2 Cu ( T e 1 − x W x ) O 6 from Néel Order to Quantum Disorder to Columnar Order*, Phys. Rev. B **98**, 064411 (2018).

[71] O. Mustonen, S. Vasala, E. Sadrollahi, K. P. Schmidt, C. Baines, H. C. Walker, I. Terasaki, F. J. Litterst, E. Baggio-Saitovitch, and M. Karppinen, *Spin-Liquid-like State in a Spin-1/2 Square-Lattice Antiferromagnet Perovskite Induced by D10–D0 Cation Mixing*, Nat. Commun. **9**, 1085 (2018).

[72] Y. Sharma et al., *Magnetic Texture in Insulating Single Crystal High Entropy Oxide Spinel Films*, ACS Appl. Mater. Interfaces 17971 (2021).

[73] B. Musicó, Q. Wright, T. Z. Ward, A. Grutter, E. Arenholz, D. Gilbert, D. Mandrus, and V. Keppens, *Tunable Magnetic Ordering through Cation Selection in Entropic Spinel Oxides*, Phys. Rev. Mater. **3**, 104416 (2019).

[74] G. H. J. Johnstone, M. U. González-Rivas, K. M. Taddei, R. Sutarto, G. A. Sawatzky, R. J. Green, M. Oudah, and A. M. Hallas, *Entropy Engineering and Tunable Magnetic Order in the Spinel High-Entropy Oxide*, J. Am. Chem. Soc. **144**, 20590 (2022).

[75] L. Su, H. Huyan, A. Sarkar, W. Gao, X. Yan, C. Addiego, R. Kruk, H. Hahn, and X. Pan, *Direct Observation of Elemental Fluctuation and Oxygen Octahedral Distortion-Dependent Charge Distribution in High Entropy Oxides*, Nat. Commun. **13**, 2358 (2022).

[76] A. Farhan, F. Stramaglia, M. Cocconcelli, N. Kuznetsov, L. Yao, A. Kleibert, C. Piamonteze, and S. Van Dijken, *Weak Ferromagnetism in Tb ( Fe 0.2 Mn 0.2 Co 0.2 Cr 0.2 Ni 0.2 ) O 3 High-Entropy Oxide Perovskite Thin Films*, Phys. Rev. B **106**, L060404 (2022).

[77] A. Farhan, M. Cocconcelli, F. Stramaglia, N. Kuznetsov, L. Flajšman, M. Wyss, L. Yao, C. Piamonteze, and S. Van Dijken, *Element-Sensitive x-Ray Absorption Spectroscopy and Magnetometry of Lu ( Fe 0.2 Mn 0.2 Co 0.2 Cr 0.2 Ni 0.2 ) O 3 High-Entropy Oxide Perovskite Thin Films*, Phys. Rev. Mater. **7**, 044402 (2023).

[78] D. I. Khomskii, *Transition Metal Compounds*, 1st ed. (Cambridge University Press, 2014).

[79] K. I. Kugel' and D. I. Khomskii, *Crystal Structure and Magnetic Properties of Substances with Or~ital Degeneracy*, (n.d.).

[80] E. Dagotto, S. Yunoki, C. Şen, G. Alvarez, and A. Moreo, *Recent Developments in the Theoretical Study of Phase Separation in Manganites and Underdoped Cuprates*, J. Phys. Condens. Matter **20**, 434224 (2008).

[81] A. R. Mazza et al., *Reversible Hydrogen-Induced Phase Transformations in La $_{0.7}$ Sr $_{0.3}$ MnO $_3$ Thin Films Characterized by In Situ Neutron Reflectometry*, ACS Appl. Mater. Interfaces **14**, 10898 (2022).

[82] T. Miao et al., *Direct Experimental Evidence of Physical Origin of Electronic Phase Separation in Manganites*, Proc. Natl. Acad. Sci. **117**, 7090 (2020).

[83] J. Yan, A. Kumar, M. Chi, M. Brahlek, T. Z. Ward, and M. A. McGuire, *Orbital Degree of Freedom in High Entropy Oxides*, Phys. Rev. Mater. **8**, 024404 (2024).

[84] A. Sarkar, D. Wang, M. V. Kante, L. Eiselt, V. Trouillet, G. Iankevich, Z. Zhao, S. S. Bhattacharya, H. Hahn, and R. Kruk, *High Entropy Approach to Engineer Strongly Correlated Functionalities in Manganites*, Adv. Mater. **35**, 2207436 (2023).

[85] R. Das, S. Pal, S. Bhattacharya, S. Chowdhury, S. K. K., M. Vasundhara, A. Gayen, and Md. M. Seikh, *Comprehensive Analysis on the Effect of Ionic Size and Size Disorder Parameter in High Entropy Stabilized Ferromagnetic Manganite Perovskite*, Phys. Rev. Mater. **7**, 024411 (2023).

[86] J. L. Braun, C. M. Rost, M. Lim, A. Giri, D. H. Olson, G. N. Kotsonis, G. Stan, D. W. Brenner, J.-P. Maria, and P. E. Hopkins, *Charge-Induced Disorder Controls the Thermal Conductivity of Entropy-Stabilized Oxides*, Adv. Mater. **30**, 1805004 (2018).

[87] Zs. Rák, J.-P. Maria, and D. W. Brenner, *Evidence for Jahn-Teller Compression in the (Mg, Co, Ni, Cu, Zn)O Entropy-Stabilized Oxide: A DFT Study*, Mater. Lett. **217**, 300 (2018).

[88] P. Mandal et al., *Orthorhombic Distortion Drives Orbital Ordering in the Antiferromagnetic 3 d 1 Mott Insulator PrTiO 3*, Phys. Rev. B **108**, 045145 (2023).

[89] Y. Tokura, *Orbital Physics in Transition-Metal Oxides*, Science **288**, 462 (2000).

[90] B. Keimer, S. A. Kivelson, M. R. Norman, S. Uchida, and J. Zaanen, *From Quantum Matter to High-Temperature Superconductivity in Copper Oxides*, Nature **518**, 179 (2015).

[91] D. I. Khomskii, *Review—Orbital Physics: Glorious Past, Bright Future*, ECS J. Solid State Sci. Technol. **11**, 054004 (2022).

[92] E. J. W. Verwey, *Electronic Conduction of Magnetite (Fe3O4) and Its Transition Point at Low Temperatures*, Nature **144**, 327 (1939).

[93] S.-W. Cheong, H. Y. Hwang, C. H. Chen, B. Batlogg, L. W. Rupp, and S. A. Carter, *Charge-Ordered States in (La,Sr ) 2 NiO 4 for Hole Concentrations n h =1/3 and 1/2*, Phys. Rev. B **49**, 7088 (1994).

[94] J. Q. Li, Y. Matsui, S. K. Park, and Y. Tokura, *Charge Ordered States in La 1 – x Sr x FeO 3*, Phys. Rev. Lett. **79**, 297 (1997).

[95] Y. Tokura, *Critical Features of Colossal Magnetoresistive Manganites*, Rep. Prog. Phys. **69**, 797 (2006).

[96] E. Dagotto, T. Hotta, and A. Moreo, *Colossal Magnetoresistant Materials: The Key Role of Phase Separation*, Phys. Rep. **344**, 1 (2001).

[97] F. Baiutti et al., *A High-Entropy Manganite in an Ordered Nanocomposite for Long-Term Application in Solid Oxide Cells*, Nat. Commun. **12**, 2660 (2021).

[98] S. Chowdhury, R. Das, K. K. Supin, M. Vasundhara, T. Bhunia, A. Gayen, and Md. M. Seikh, *Effect of Local Chemical Disordering on Magnetic Properties in High Entropy Manganite of Variable Hole Concentration*, Ceram. Int. S0272884223042669 (2023).

[99] Z. Shi, J. Zhang, J. Wei, X. Hou, S. Cao, S. Tong, S. Liu, X. Li, and Y. Zhang, *A-Site Deficiency Improved the Thermoelectric Performance of High-Entropy Perovskite Manganite-Based Ceramics*, J. Mater. Chem. C **10**, 15582 (2022).

[100] R. Das, S. Bhattacharya, S. Chowdhury, S. Sen, T. K. Mandal, T. Bhunia, A. Gayen, M. Vasundhara, and Md. M. Seikh, *High Entropy Effect on Double Exchange Interaction and Charge Ordering in Half Doped Nd0.5Sr0.5MnO3 Manganite*, J. Alloys Compd. **951**, 169950 (2023).

[101] I. I. Mazin, D. I. Khomskii, R. Lengsdorf, J. A. Alonso, W. G. Marshall, R. M. Ibberson, A. Podlesnyak, M. J. Martínez-Lope, and M. M. Abd-Elmeguid, *Charge Ordering as Alternative to Jahn-Teller Distortion*, Phys. Rev. Lett. **98**, 176406 (2007).

[102] H. Park, A. J. Millis, and C. A. Marianetti, *Site-Selective Mott Transition in Rare-Earth-Element Nickelates*, Phys. Rev. Lett. **109**, 156402 (2012).

[103] S. Johnston, A. Mukherjee, I. Elfimov, M. Berciu, and G. A. Sawatzky, *Charge Disproportionation without Charge Transfer in the Rare-Earth-Element Nickelates as a Possible Mechanism for the Metal-Insulator Transition*, Phys. Rev. Lett. **112**, 106404 (2014).

[104] M. A. Buckingham, B. Ward-O'Brien, W. Xiao, Y. Li, J. Qu, and D. J. Lewis, *High Entropy Metal Chalcogenides: Synthesis, Properties, Applications and Future Directions*, Chem. Commun. **58**, 8025 (2022).

[105] B. Jiang et al., *High-Entropy-Stabilized Chalcogenides with High Thermoelectric Performance*, Science **371**, 830 (2021).

[106] Y. Sharma et al., *High Entropy Oxide Relaxor Ferroelectrics*, ACS Appl. Mater. Interfaces **14**, 11962 (2022).

[107] H. Qi, L. Chen, S. Deng, and J. Chen, *High-Entropy Ferroelectric Materials*, Nat. Rev. Mater. (2023).

[108] B. Liu, W. Yang, G. Xiao, Q. Zhu, S. Song, G.-H. Cao, and Z. Ren, *High-Entropy Silicide Superconductors with W 5 Si 3 -Type Structure*, Phys. Rev. Mater. **7**, 014805 (2023).

[109] K. Wang et al., *Structural and Physical Properties of High-Entropy REBa2Cu3O7-δ Oxide Superconductors*, J. Supercond. Nov. Magn. **34**, 1379 (2021).